\RequirePackage{fix-cm}
\documentclass[twocolumn,natbib]{svjour3}                     
\usepackage[breaklinks]{hyperref}
\usepackage{graphicx}
\usepackage{textcomp}
\usepackage{amsmath}
\usepackage{amssymb}
\usepackage{booktabs}
\usepackage{xcolor}
\usepackage{textcomp}
\smartqed 
\usepackage{latexsym}
\usepackage{etoolbox}

\newcommand{\splitcell}[2][c]{%
  \begin{tabular}[c]{@{}c@{}}\strut#2\strut\end{tabular}%
}
\begin{document}

\title{Tomographic long-distance \textmu PIV to investigate the small scales of turbulence in a jet at high Reynolds number}



\author{Daniele Fiscaletti \and Daniele Ragni \and Edwin F.J. Overmars \and Jerry Westerweel \and Gerrit E. Elsinga}


\institute{Daniele Fiscaletti \at
              Aerodynamics, Wind Energy, Flight Performance and Propulsion Department, Faculty of Aerospace, TU Delft,  Kluyverweg 1, 2629HS, Delft, The Netherlands \\
              \email{d.fiscaletti@tudelft.nl}}
              
\date{Received: date / Accepted: date}

\maketitle

\begin{abstract}
The small scales of turbulence in a high-Rey\-nolds-number jet ($Re_{\lambda} \approx 350$) are investigated with a \textmu PIV setup to overcome the optical limitations of conventional tomographic PIV setups. With the aim of validating the performances of tomographic long-distance
\textmu PIV, analyses are carried out involving statistical
aspects of the small scales of turbulence. 
The technique is assessed and the data are bench-marked to be applied to the analysis of any three dimensional small-scale phenomena in large-scale flow domains.
\end{abstract}

\section{Introduction and problem definition}
\label{introduction}

Turbulence involves fluid motions over a wide range of length scales. While classical theory assumed that the small scales were largely independent of the large scales, it is becoming increasingly clear that direct interactions between these disparate scales are important \citep{Shen2000,LaPorta2001,Biferale2012,Fiscaletti2016}. Moreover, the magnitude of the extreme velocity gradients and dissipation are strongly Reynolds number dependent \citep{Buaria2019,Elsinga2017,Elsinga2020}, which suggests large-scale influences at small scales. These influences remain to be fully understood. Therefore, experimental investigations of small-scale turbulence at high Reynolds number are of great interest. \\
Small-scale turbulence is essentially three-dimensional and is closely associated with velocity gradients. Flows at high Reynolds numbers present an effective separation between the energetic large-scale motions and the small dissipative scales, as is required to elucidate their interaction. Furthermore, high Reynolds number turbulence occurs in many industrial and natural flows, e.g. flow over a wind turbine blade and atmospheric/oceanic flow, which makes it relevant from an application point of view. 
However, 3D measurements of the small scales are extremely challenging due to the spatial resolution requirements, especially in laboratory flows at high Reynolds numbers. So far, the experimental study of the small scales in high Reynolds numbers turbulence was mainly performed using multi-probe hot-wire anemometry \citep{Wallace2010}. Nano-scale thermal anemometry probes were also manufactured to reduce the spatial filtering and the intrusivity of the probes, which led to extremely accurate turbulence measurements \citep{Vallikivi2011}. However, regardless of the size of the probes, aspects involving the three-dimensional spatial organization of the small-scale structures cannot be investigated using hot-wire anemometry. Typical small-scale flow structures include vortex tubes and dissipation sheets, whose linear core size is approximately 10$\eta$ \citep{Siggia1981,Jimenez1993,Ganapathisubramani2008,Elsinga2017}. Here, $\eta$ is the global Kolmogorov length scale. However, it should be noted that locally the Kolmogorov length scale can be smaller within highly turbulent regions of the flow, as shown by \cite{Buaria2019} and \cite{Elsinga2020}, which implies that smaller structures exist. \\
\cite{Tokgoz2012} have shown that a cross-correlation window size of 8$\eta$ is needed at $75\%$ window overlap in order to resolve the viscous dissipation using tomographic particle image velocimetry (PIV), which is consistent with the mentioned structure size. Taking the present laboratory jet (section \ref{sec:flow}) as an example, the required window size is approximately 0.5 mm, or less when the Reynolds number is increased. This spatial resolution requirement goes beyond the 1-2 mm, which is commonly obtained by tomographic PIV \citep{Elsinga2006, Scarano2009, Jodai2016,Debue2021}. Higher spatial resolution is achieved at short distances when using a microscopic objective \citep{Kim2011,Kim2012}. Digital holographic PIV is capable of very high spatial resolution in all directions, but its working distance is fundamentally limited by the sensor size. Therefore, it is suited mostly for near wall turbulence \citep{Sheng2009,Willert2017}. In scanning PIV \citep{Casey2013,Lawson2014} and cinematographic PIV, which relies on the Taylor hypothesis \citep{vanDoorne2007,Ganapathisubramani2007}, the spatial resolution is limited in the third direction by the light sheet thickness, which is difficult to decrease below 0.5 mm in large facilities. At present, these technical and physical limitations prevent the holographic and scanning approaches from achieving high spatial resolution at large working distances. \\
Here, we explore the extension of tomographic PIV approaches with long-range microscopic objectives in order to achieve high spatial resolution at large working distances. The technique will be referred to as tomographic long-range \textmu PIV. As explained above, this development opens up new opportunities in the investigation of turbulence at high Reynolds numbers. \\
Long-range microscopic objectives have been used successfully in planar PIV to resolve small-scale turbulence in many different turbulent flows, such as a boundary layer, a pipe flow, a jet, and a flow in an internal combustion chamber \citep{Kahler2006,Lindken2002,Fiscaletti2014b,Ma2017}. However, only two velocity components and one vorticity component were obtained in these experiments. The present extension to tomographic long-range \textmu PIV enables to capture all three velocity components and all nine velocity gradients. \\
The paper is structured as follows. Section \ref{sec:flow} describes the jet flow under investigation and its turbulence properties, which are characterized through hot-wire anemometry and Pitot tube measurements. In Section \ref{sec:setup}, the limitations of studying high-Reynolds-number turbulence using pre-assembled microscopes are discussed, and an alternative experimental approach that overcomes these limitations is proposed, also illustrating the effects of the design parameters on the measurement. Section \ref{sec:processing} is devoted to the processing of the data and to the a-posteriori attenuation of the measurement noise. In Section \ref{sec:results}, the experimental method is validated based on the analysis of the statistical properties of the small-scale turbulence. The main findings are then summarized in Section \ref{sec:conclusions}. 
\section{The turbulent flow}
\label{sec:flow} 
\begin{figure}
  \includegraphics[width=0.48\textwidth]{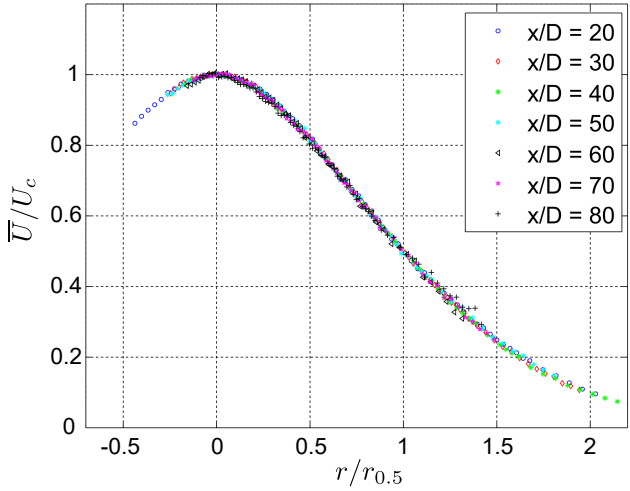}
\caption{Mean streamwise velocity against radial position at different nozzle diameters \textit{D} downstream from the jet nozzle, from measurements with hot-wire anemometry.}
\label{mean_vel}
\end{figure}

\begin{figure}
  \includegraphics[width=0.48\textwidth]{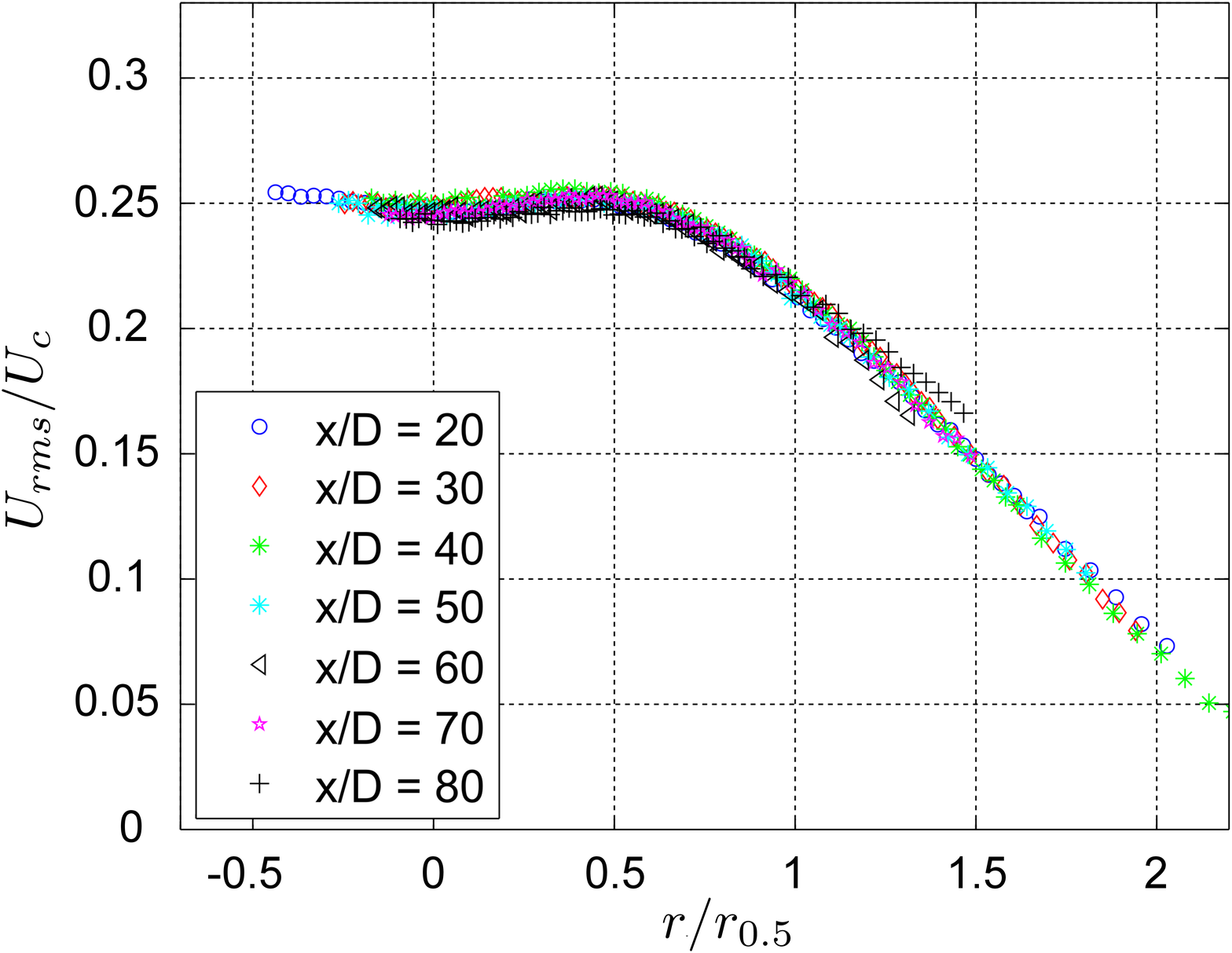}
\caption{Root-mean-square (r.m.s.) streamwise velocity against radial position at different nozzle diameters \textit{D} downstream from the jet nozzle, from measurements with hot-wire anemometry.}
\label{rms_vel}
\end{figure}

Experiments were carried out within a fully-developed region of a turbulent jet of air. A nozzle with a diameter of $D =$ 8$\times$10$^{-3}$ m creates a jet with a mean velocity of $U_{J} = 125$ ms$^{-1}$, obtained from the dynamic pressure measured with a Pitot tube. The non-dimensional numbers at the nozzle are $Re_{D} = 6.6 \times 10^{4}$ and $Ma = 0.37$. While the jet is weakly compressible near the nozzle, the measurement location was chosen at about 70 diameters distance, where the Mach number reduces to $0.03$. A schematic of the experimental apparatus producing the turbulent air jet under investigation is shown in figure 3 of \cite{Fiscaletti2014a}. Further details on the settling chamber and the shape of the nozzle are given by \cite{Slot2009}. The jet flow was characterized with hot-wire anemometry (HWA) in a constant temperature configuration. The HWA velocity profiles are additionally used to benchmark the results obtained from tomographic PIV. A system of coordinates is introduced, with the \textit{xy}-plane containing the laser sheet of the PIV measurements. The HWA measurements were taken with a Dantec 55P11 sensor. The overheat ratio was set to 0.7 to ensure a constant wire temperature of approximately $220^{\circ}$C. The feedback control of the sensor temperature was adjusted through a Dantec Dynamics 56C17 Wheatstone bridge. From the measurements with hot-wire anemometry we could estimate the turbulent length scales at different downstream distances from the nozzle.
Firstly, the mean dissipation rate in the jets is estimated with \citep{Panchapakesan1993}:
\begin{equation}
    \bar{\varepsilon} \approx 0.015 \frac{U_{c}^{3}}{r_{0.5}}
\end{equation}
where $U_{c}$ is the center-line velocity, and $r_{0.5}$ is the jet half-width. Based on the local dissipation rate, the Kolmogorov length scale is calculated according to:
\begin{equation}
    \eta = \Bigg(\frac{\nu^{3}}{\bar{\varepsilon}}\Bigg)^{\frac{1}{4}}
\end{equation}
where $\nu$ is the kinematic viscosity. The radial profiles of the mean and the root-mean-square (r.m.s.) velocities, respectively expressed by $\overline{U}$ and $U_{rms}$, were scaled with the mean center-line velocity $U_{c}$ of the flow. These velocity profiles collapsed on a single curve after scaling the radial displacement $r$ by $r_{0.5}$, thus showing the self-similar behavior typical of the jet flows in the fully-developed region, i.e. $x/D>20$ (Figs. \ref{mean_vel} and \ref{rms_vel}).
The described measurements enabled to infer the macroscopic characteristics of the jet, such as the spreading rate and the decay rate of the center-line velocity (see table \ref{flow_turbulence}). Both quantities exhibited a good agreement with other studies available in the literature \citep{Panchapakesan1993,Hussein1994}. The maximum deviation for both the spreading rate and the decay rate is within $5\%$, when compared with the study by \cite{Hussein1994}. Based on the acquired velocity statistics and under the hypothesis of isotropic turbulence, the Taylor micro-scale was estimated using:
\begin{equation}
    \lambda =  u_{rms}\sqrt{15\frac{\nu}{\bar{\varepsilon}}}
\end{equation}
where $u_{rms}$ is the r.m.s. of the axial velocity fluctuations.\\
The \textmu PIV measurements were taken along the center-line at a downstream distance of $x/D = 70$ from the nozzle. At this location, the most relevant turbulent quantities were determined as $\eta = 57$ \textmu m; $\lambda = 2.19$ mm, and
$Re_{\lambda} = u_{rms}\lambda/\nu = 367$. The Reynolds number of the turbulent flow is sufficiently large to investigate both small-scale motions and the vorticity stretching motions, as these can be considered to be fully developed beyond a Reynolds number of 250 \citep{Elsinga2017}. A detailed summary of the estimated flow characteristics can be found in table \ref{flow_turbulence}.
\begin{table}[h]
    \begin{tabular}{l r}
    \toprule
Jet exit diameter $D$ [m]   & 8$\times$10$^{-3}$   \\
Jet exit velocity  ($U_{J}$) [ms$^{-1}$]  &  124.8  \\
Reynolds number based on $D$ ($Re_{D}$) & 6.6$\times$10$^{4}$   \\
Spreading rate ($S=\frac{dr_{0.5}(x)}{dx}$) & 0.096  \\
$S$ from \cite{Hussein1994} &  0.102  \\
$S$ from \cite{Panchapakesan1993} &  0.096  \\
Velocity decay ($B=\frac{U_{0}(x)(x-x_{0})}{U_{J}D}$) & 5.6  \\
$B$ from \cite{Hussein1994} &  5.9  \\
$B$ from \cite{Panchapakesan1993} &  6.1 \\
Measurement location ($x/D$)  &  70  \\
Centerline velocity $U_{c}$ [ms$^{-1}$] & 10.56 \\
Jet half width $r_{0.5}$ [mm] & 52.2 \\
Taylor microscale $\lambda$ [mm] & 2.19  \\
Dissipation rate $\bar{\varepsilon}$ $[m^{2}s^{-3}]$ & 337.6  \\
Kolmogorov length scale $\eta$ based on $\bar{\varepsilon}$ [\textmu m]  & 57  \\
Kolmogorov time scale $\tau_{\eta}$ based on $\bar{\varepsilon}$ [\textmu s] & 213 \\
$\lambda/\eta$ ratio & 38.4 \\
Reynolds number based on $\lambda$  $Re_{\lambda}$ & 367  \\
\bottomrule
    \end{tabular}
    \caption{Flow characteristics as estimated from hot-wire anemometry and Pitot tube measurements.}
    \label{flow_turbulence}
\end{table}

\section{Experimental setup}
\label{sec:setup}

\begin{figure}
  \includegraphics[width=0.48\textwidth]{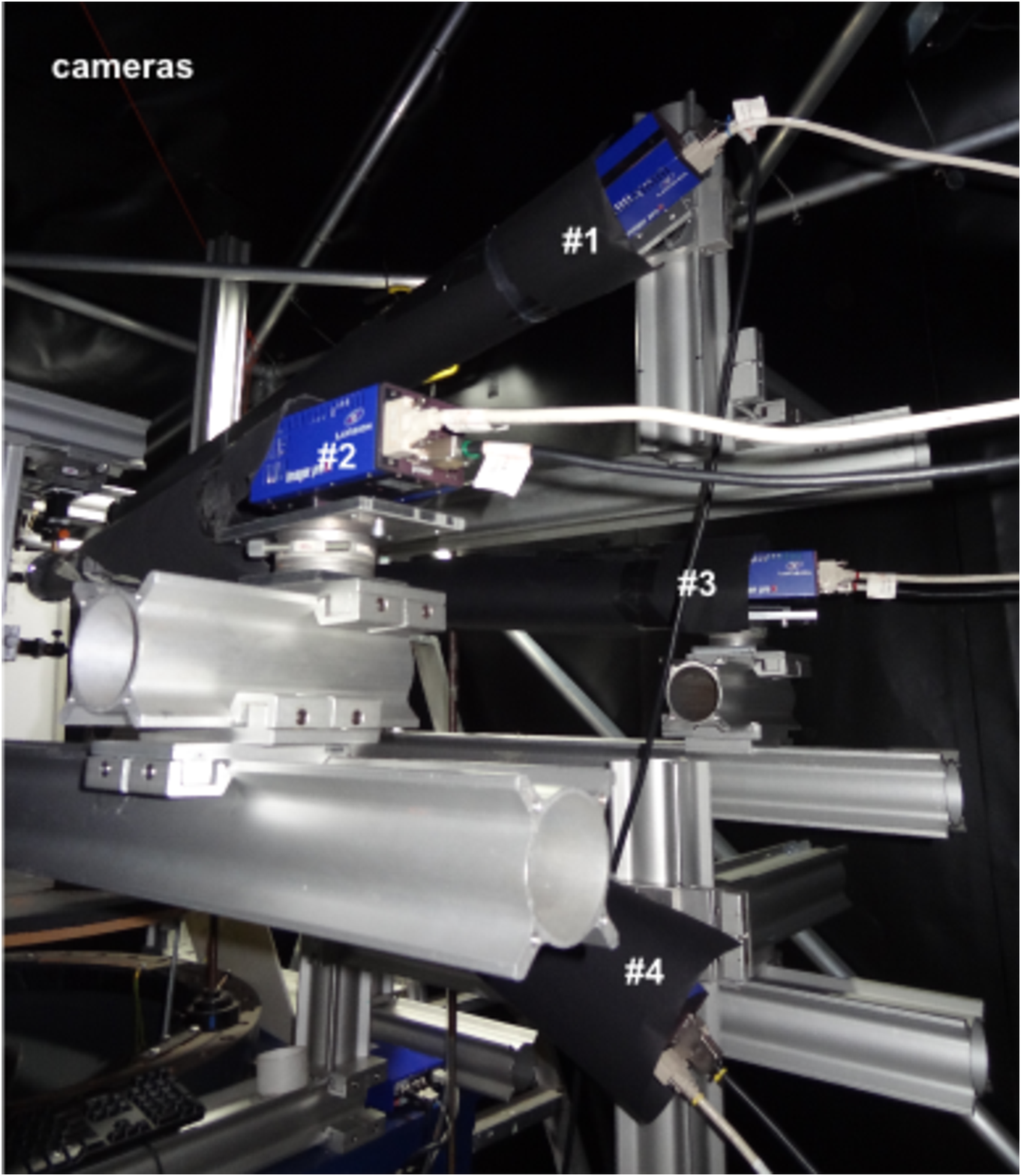}
\caption{Photo of the experimental setup showing the positioning of the four cameras in the 3D space. For each camera, the optical path between the CCD sensor and the associated lens was shielded to reduce the capturing of day light.}
\label{foto_real_1}
\end{figure}

\begin{figure}
  \includegraphics[width=0.48\textwidth]{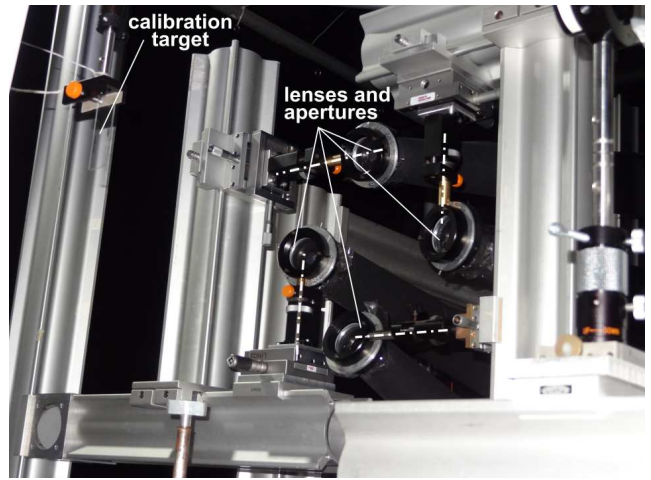}
\caption{Photo of the experimental setup showing the calibration target and the positioning of the four lenses and of the related four diaphragms. Dotted-dashed lines highlight the axes of rotation of the four lenses. The Scheimpflug condition is achieved by rotating the lenses around these axes. }
\label{foto_real_2} 
\end{figure}

\begin{figure}[ht]
  \includegraphics[width=0.48\textwidth]{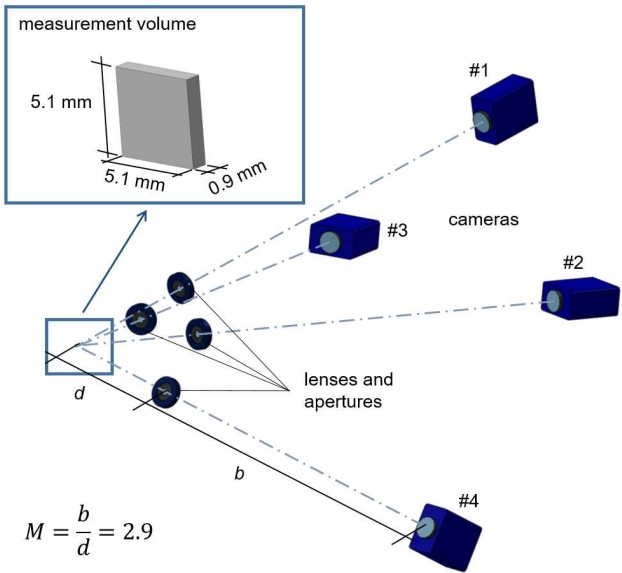}
\caption{Schematic to scale of the experimental setup showing the mutual positioning of the cameras, the optics, and the measurement plane.}
\label{sketch}
\end{figure}

\begin{figure*}
\centering
  \includegraphics[width=0.8\textwidth]{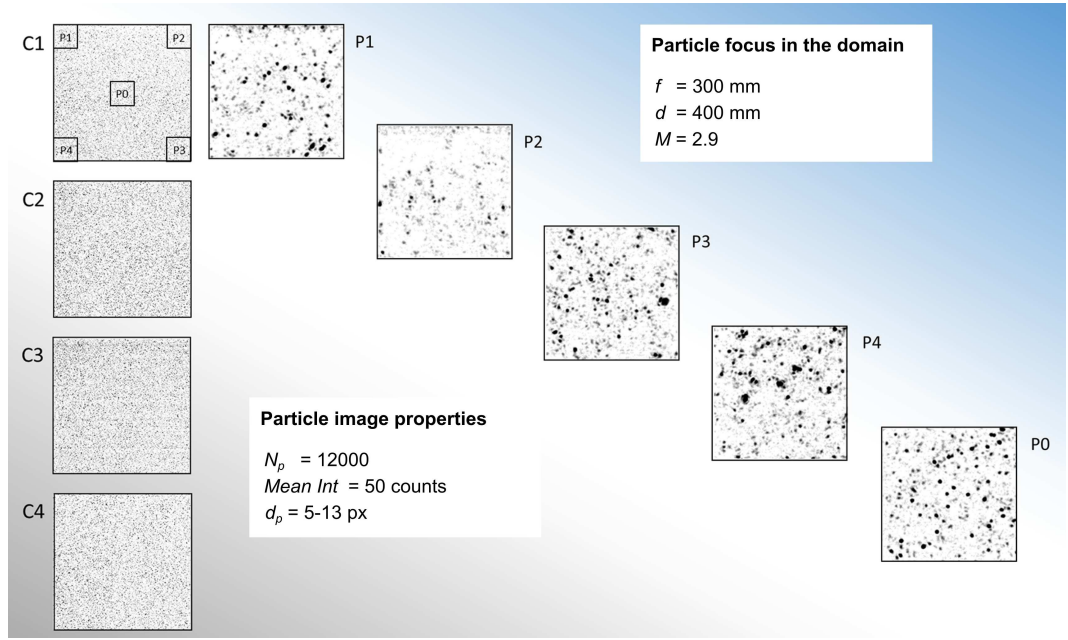}
\caption{PIV images from the four different cameras, i.e. C1, C2, C3, and C4. The five zoomed views show the quality and the homogeneity of the particle images in five different regions of the PIV image from camera $\#1$, i.e., in P1, P2, P3, P4, and P5.  }
\label{particle_image} 
\end{figure*}

Conventional long-distance microscopes such as Questar QM-1 and Infinity K2 have been typically employed in experiments of long-range \textmu PIV \citep{Kahler2006,Eichler2012,Ma2017}. However, when extending long-range \textmu PIV into a three-dimensional system the cameras should  be positioned at an opening angle in order to capture the out-of-plane component of the velocity vector. If conventional pre-assembled long-distance microscopes are employed, the Scheimpflug condition cannot be verified, which limits the field of view when images are in focus, see figure 2 of \cite{Fiscaletti2014b}. 
To overcome these limitations a new experimental approach was employed in this study. Single plano-convex lenses were positioned on the optical path between the camera and the measurement object. Despite the complexity in properly aligning the lenses with the camera sensors, the present configuration allows for a small rotation of the lens around an axis orthogonal to the optical path, in order to satisfy  the Scheimpflug condition, which greatly enhances the particle focus throughout the image. The described experimental approach is also advantageous from an economical point of view, as the total cost of the optics is reduced by two orders of magnitude compared with an optical setup with long-distance microscopes. The measurement system is presented in figures \ref{foto_real_1} and \ref{foto_real_2}. The capturing of day light is reduced by shielding the four optical paths between the cameras and the lenses by means of paper cones, as shown in figure \ref{foto_real_1}. With the aim of controlling the depth of focus and modulating the incoming light intensity, apertures were positioned right behind the lenses, as shown in figure \ref{foto_real_2}. For the present study we aim at a magnification factor of around 2.5 at a distance of approximately 500 mm, analogous to the experiment of long-distance \textmu PIV by \cite{Fiscaletti2014a}. The required focal length, \textit{f}, follows from the desired magnification, \textit{M}, and the stand off distance, i.e. the distance between the lens and the measurement plane, \textit{d}, which is determined by the size of the flow facility. From the lens equation and the definition of the magnification, it follows that:
\begin{equation}
    f = \frac{d \cdot M}{(M+1)}
\end{equation}
leading to a focal length of $f=300$ mm. The aperture diameter $D_{l}$ is estimated from the expression for the focal depth $\delta z$:
\begin{equation}
\label{eqn:depth_focus}
    \delta z = 4 \cdot \lambda \cdot f_{\#}^{2} \cdot \Big(1 + \frac{1}{M} \Big)^2
\end{equation}
where $f_{\#}=f/D_{l}$ and $\lambda = 532$ nm is the wavelength of the laser light. 
Equation \ref{eqn:depth_focus} can be solved for $D_{l}$ when $\delta z$ is prescribed to be larger than the light sheet thickness. For a depth of focus of 1 mm, equivalent to approximately $15\eta$, $D_{l}$ is 19 mm. A schematic drawn to scale of the experimental setup is presented in figure \ref{sketch}. \\
 Samples of PIV images obtained from the four cameras are shown in figure \ref{particle_image}. In addition, five zoomed views from camera $\#1$ show the quality of the particle images, and their homogeneity throughout each PIV image. The particle images appear in focus independently of the region where they are located. Besides the described array of four lenses, the \textmu PIV system consisted of a Nd:YAG laser (Quanta-Ray, Spectra-Physics), four CCD-cameras with a $2,048 \times 2,048$ pixel format sensor (pixel size, $d_{r} = 6.45$ \textmu m). PIV seeding consisted of DEHS droplets [Di(2-ethylhexyl) sebacate, sebacic acid], generated from a Laskin nozzle generating particle tracers with a median peak for a size of $1$ \textmu m \citep{Ragni2011}. Based on a particle diameter of $1$ \textmu m, the response time of the seeding droplets is computed to be 1 \textmu s, as shown by \cite{Ragni2011}. On the other hand, the Kolmogorov time scale at the measurement location is $\tau_{\eta} = \sqrt{\nu / \bar{\varepsilon}} = 200$ \textmu s, therefore two orders of magnitude larger than the time response of the particles. This indicates that the velocity fluctuations in the flow can be followed accurately by these particles. For the calibration, a glass target with a grid spacing of $10$ \textmu m was employed. The traversing of the target during the calibration process was carried out with a remote linear traversing  actuator (Zaber\textsuperscript{TM}, Edmund Optics). The spatial resolution of the traversing is of $4.8 \cdot 10^{-2}$ \textmu m, with a (unidirectional) accuracy from the specifications of the manufacturer of $15$ \textmu m. Five planes along the viewing direction separated by a distance of 0.6 mm were acquired for the tomographic calibration procedure.
\section{Data processing and assimilation}
\label{sec:processing}
\begin{figure}[ht]
  \includegraphics[width=0.46\textwidth]{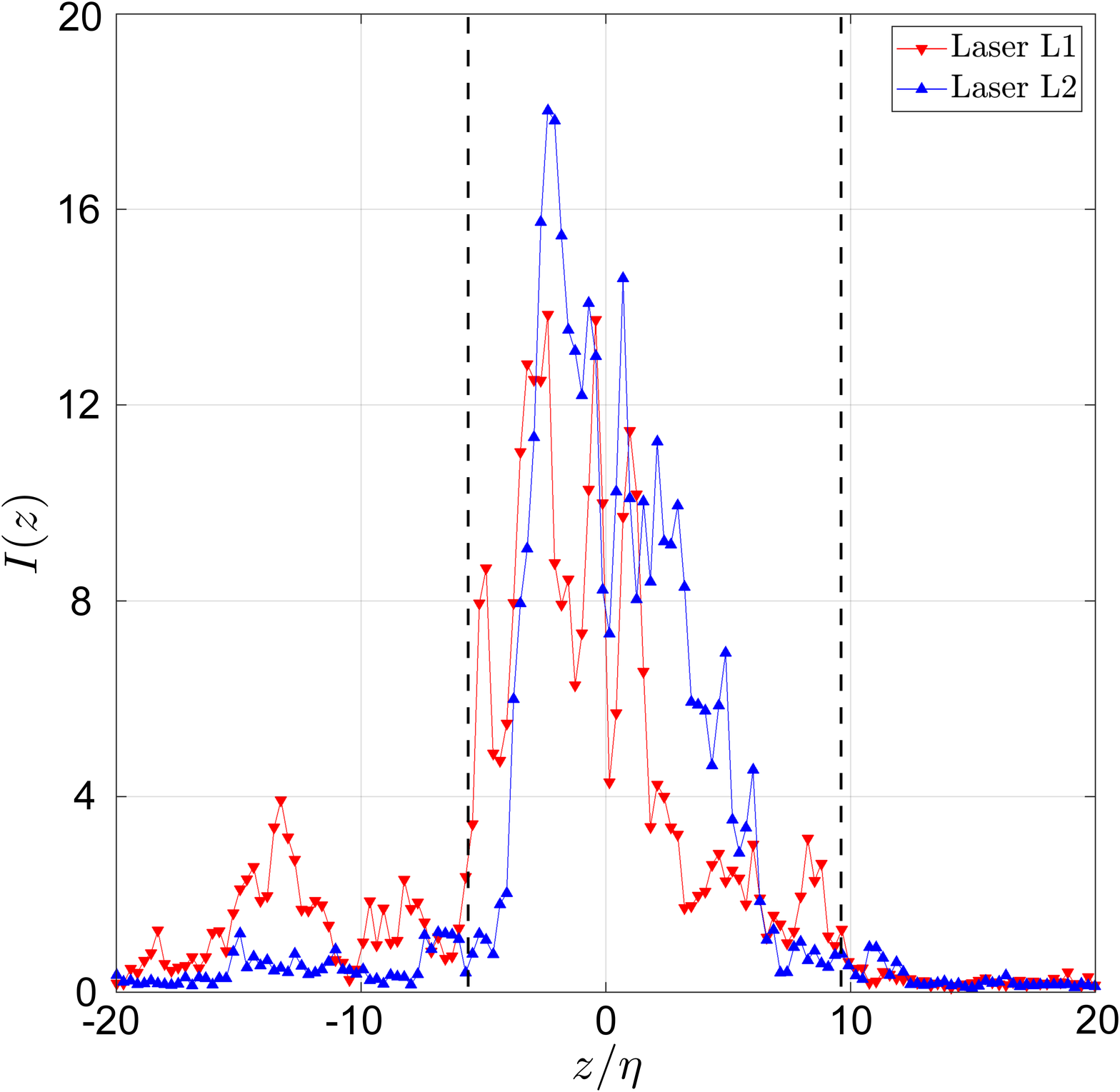}
\caption{Light intensity distribution in a tomographic reconstruction  averaged along the x and y directions. L1 corresponds to the first image (first laser pulse) and L2 to the second one (second laser pulse). The thickness of the light sheet can be estimated to be approximately $16\eta$, equivalent to 912 \textmu m. The dashed lines identify the extent along the \textit{z}-axis of the reconstructed volume for cross-correlation.}
\label{laser_sheet}
\end{figure}

The acquired PIV images were processed with the software DaVis 8.4 from LaVision, using the Fast MART algorithm for the reconstruction of the three-dimensional particle intensity distribution. However, first a calibration process was carried out by fitting a pinhole camera model to the calibration plate images, further refined by an iterative self-calibration process \citep{Wieneke2008}. Three self-calibration iterations were performed until the disparity values in the imaged domain were reduced to $<$0.01 px ($\approx0.034$ \textmu m). From the pinhole model, the orientation angles of the four cameras together with their distance to the target and the computed focal length in the 3D domain could be determined and verified. The orientation angles of the cameras are presented in table \ref{angles}. The geometrical opening angle between cameras $\#1$ and $\#4$ and between cameras $\#2$ and $\#3$ are both approximately 40$^{\circ}$. 
\begin{table}[h]
\begin{centering}
    \begin{tabular}{c c c c} 
    \toprule
\null\hfill camera \hfill\null &  \textit{x}-axis $[^{\circ}]$  & \textit{y}-axis $[^{\circ}]$ & \textit{z}-axis $[^{\circ}]$ \\
$\#1$    & +7 & +25 & -90  \\
$\#2$    & +18 & +4 & +1  \\
$\#3$    & -10 & -1 & +1   \\
$\#4$    & +7 & -16 & -90  \\
\bottomrule
    \end{tabular}
    \caption{Inclination angles of the four cameras along each of the three Cartesian axes with respect to a direction orthogonal to the plane \textit{xy}, according to the tomographic reconstruction algorithm.}
    \label{angles}
    \end{centering}
\end{table}
From the tomographic reconstruction the light intensity distribution of the two laser pulses was estimated by averaging the light intensity distribution along the \textit{x}- and \textit{y}-axis. Figure \ref{laser_sheet} shows the light intensity distribution obtained from a single pair of tomographic reconstructions. The light pulses present a fairly good overlap. Furthermore, the ratio of reconstructed intensity inside the light sheet relative to reconstruction noise outside the sheet (signal-to-noise ratio) is about 12 at the center of the reconstructed volume, although this value reduces when the edges of the reconstructed volume are approached. The light intensity distribution of these laser pulses is also used to determine the thickness of the light sheet, which is of approximately $16\eta$, equivalent to 912 \textmu m. A depth of focus of 1000 \textmu m ensures that all the particles in the laser domain are properly imaged in focus (see figure \ref{particle_image}). The size of the reconstruction volume before cross-correlation was of $1500 \times 1500 \times 270$ px$^{3}$ equivalent to $5.1 \times 5.1 \times 0.9$ mm$^{3}$, which expressed in Kolmogorov length scale corresponds to $89 \times 89 \times 16$ $\eta^{3}$. The extent of the reconstructed volume along the \textit{z}-axis is identified in figure \ref{laser_sheet} using dashed lines. 
 Velocity vectors were obtained by a multi-grid correlation iterative process starting from an initial uniform correlation volume of size $128 \times 128 \times 128$ vx down to $96 \times 96 \times 96$ vx, equivalent to $330 \times 330 \times 330$ \textmu m$^{3}$, which in Kolmogorov length scale corresponds to $5.8 \times 5.8 \times 5.8$ $\eta^{3}$. From applying a window overlap of $75\%$, a vector spacing of approximately $80$ \textmu m was obtained in each direction, corresponding to $1.5$ $\eta$. The vector fields were additionally processed with a median outlier removal filter re-interpolating less than $5\%$ of the full amount of vectors \citep{Westerweel2005}. A summary of the optical and reconstruction parameters of the tomographic \textmu PIV measurement is reported in table \ref{reconstruction}. \\
 \begin{table}[h]
\begin{centering}
    \begin{tabular}{l r}
    \toprule
Focal length $f$ [mm]  & 300 \\
$f_{\#}$ ($f/D_{l}$) &  16 \\
Depth of focus $\delta$\textit{z} [\textmu m] & 1000\\
Light sheet thickness [\textmu m] & 900 \\
Field of view [mm$^{2}$] & $5.1 \times 5.1$\\
Volume size [vx] & $1500 \times 1500 \times 270$\\
Scaling factor [px mm$^{-1}$] & 294\\
Correlation volume [vx] & $96 \times 96 \times 96$\\
Spatial resolution [\textmu m] & 330 ($6\eta$)\\
Vector spacing [\textmu m] & 83 ($1.5\eta$)\\
Volume overlap [$\%$] & 75\\
Object distance [mm] & 400\\
Magnification factor \textit{M} & 2.9\\
Pixel pitch [\textmu m] & 7.4\\
Particle image diameter [\textmu m] & 72\\
Particle image diameter [px] & 9.8\\
\bottomrule
    \end{tabular}
    \caption{Optical and reconstruction parameters of the tomographic \textmu PIV measurement.}
    \label{reconstruction}
    \end{centering}
\end{table}
 The obtained vector fields were subsequently processed with a finite time marching algorithm based on a volume-in-cell (VIC) technique \citep{Schneiders2014,Schneiders2016b}. The main aim of this processing was to impose a divergence-free condition to the reduce measurement noise in the three directions. Dealing with divergence-free velocity fields is necessary to relate vorticity to velocity at each integration time step, which is at the basis of applying VIC algorithms. A reduction of the divergence error in volumetric measurements leads to improved flow statistics as shown by \cite{DeSilva2013b} and \cite{Schiavazzi2013}. 
In the VIC method, the vorticity field is discretized by a number of Lagrangian vortex particles on a Cartesian grid. The obtained vortex particles are governed by two ordinary differential equations, a first equation for advection, and a second equation aimed at updating the vorticity strength of each particle, which accounts for vortex stretching. Solving these differential equations gives the configuration of the vortex particles at the subsequent time step. From the updated configuration of the vortex particles, the vorticity field is calculated on the grid through an interpolation procedure, while the application of the Poisson equation enables to determine the velocity vector field associated to the new configuration, therefore at the new time step. Further details on the VIC algorithm can be found in \cite{Schneiders2014}.\\
 The VIC algorithm was recursively applied to create a time series for which an adjoint-method iteration of the vorticity is carried out with the subsequent VIC+ method \citep{Schneiders2016}. The iterative approach for the vorticity enables to re-interpolate the velocity information on the same original grid. Although the method solves the incompressible Navier-Stokes equations in the form of the vorticity transport equations, inputs from the solver are the originally reconstructed velocity fields. 
 The VIC+ upgrade builds upon the original formulation, but with the vorticity field computed from an adjoint optimization. The VIC+ calculates both the velocity and the velocity material derivative on the same grid points as those from the original Cartesian grid. These physical quantities are then plugged in a cost function comparing the measured versus the estimated values, and an iterative process makes the value of this cost function decrease below a convergence threshold. Further details on the VIC+ method can be found in \cite{Schneiders2016}. The application of the VIC+ algorithm to a time series created by recursively running the VIC algorithm on a single-snapshot PIV has never been reported to date. \cite{Schneiders2016}, however, showed that VIC+ leads to a higher accuracy of the results when applied to experimental time-resolved velocity data. Recently, time series obtained from a VIC marching scheme were favoura\-bly compared against time-resolved experimental measurements by \cite{Schneiders2016b,Schneiders2018}. From these observations, the use of VIC+ on a time-dependent dataset generated using a VIC marching scheme is expected to improve the accuracy of the results, and it is therefore employed in the present study.  \\
 By recursively running the VIC algorithm for five time steps, five time-resolved velocity vector fields were obtained with a finite marching with $\Delta t=2$ \textmu s, verifying the stability conditions for the marching solver with the current vector spatial resolution. 
From the obtained dataset, the velocity material derivative was estimated and it was applied for the central time-step VIC+. It is worth noting that the contribution of advection is dominant, and that the flow over these time series is comparable to frozen turbulence, given that the time step $\Delta t$ is two orders of magnitude smaller than the Kolmogorov time scale (see table \ref{flow_turbulence}).  Fifty iterations were run, with the cost function of VIC+ converging till the residual $J$ values are below 0.8. The main benefits of using the described treatment of the velocity fields from tomographic \textmu PIV lies in the improvement of resolution details of the turbulence structures \citep{Schneiders2016}. However, in this specific study, for consistency with the original data, an increase of resolution was not carried out.   

 \section{Assessment of the experimental dataset}
\label{sec:results}
\begin{figure*}[ht]
 \includegraphics[width=0.98\textwidth]{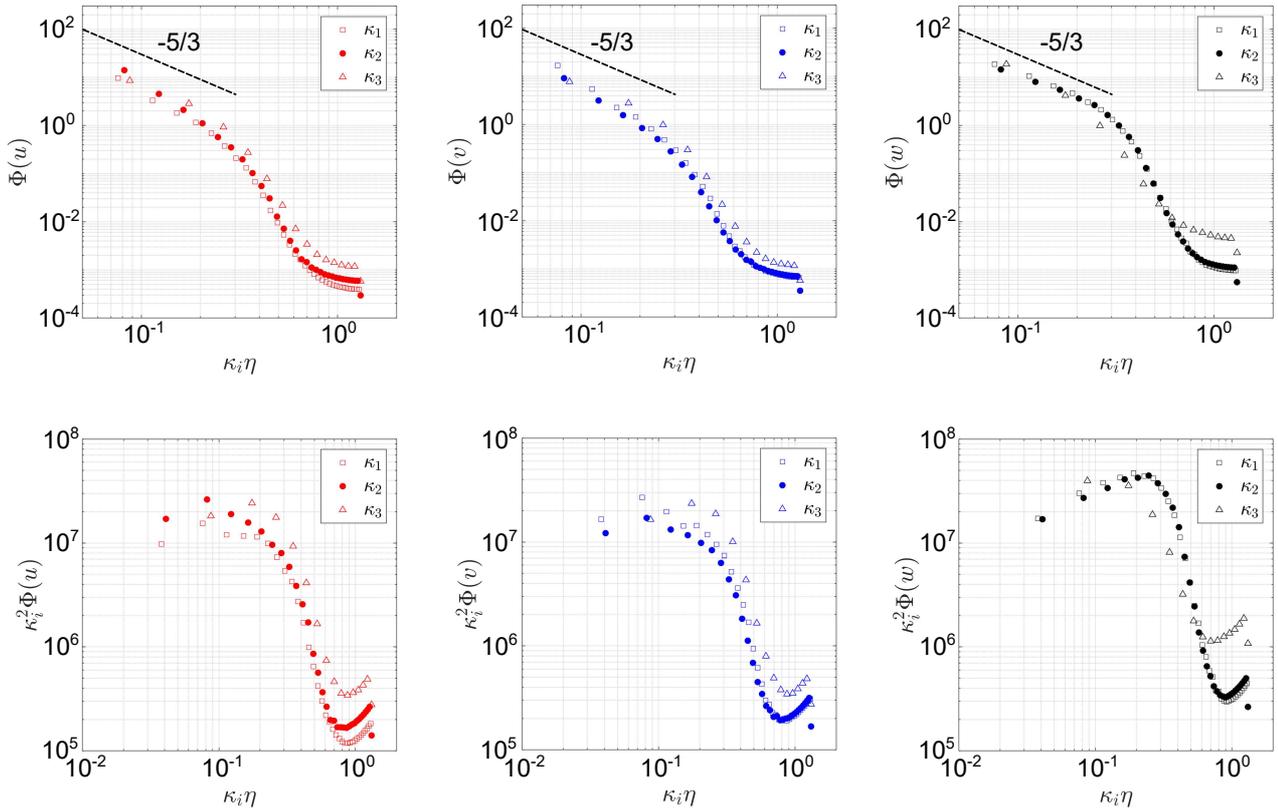}
\caption{Spectra of the streamwise (left), transversal (middle), and out-of-plane components of the velocity (right column) along the three Cartesian axes; \textit{top}, velocity spectra and, \textit{bottom}, dissipation spectra.  }
\label{spectra}       
\end{figure*}

\begin{figure}[hb]
  \includegraphics[width=0.48\textwidth]{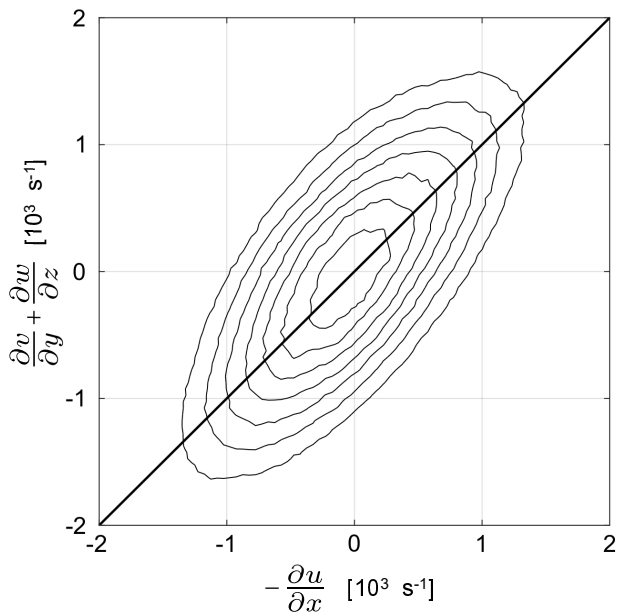}
\caption{Joint probability density function of $- \partial u / \partial x$ and $( \partial v / \partial y + \partial w / \partial z)$ with VIC+ and the regression filter. Contours in range from $7 \cdot 10^{-8}$ to $2.5 \cdot 10^{-7}$ in increments of $3 \cdot 10^{-8}$.}
\label{continuity}       
\end{figure}

\begin{figure}[hb]
  \includegraphics[width=0.48\textwidth]{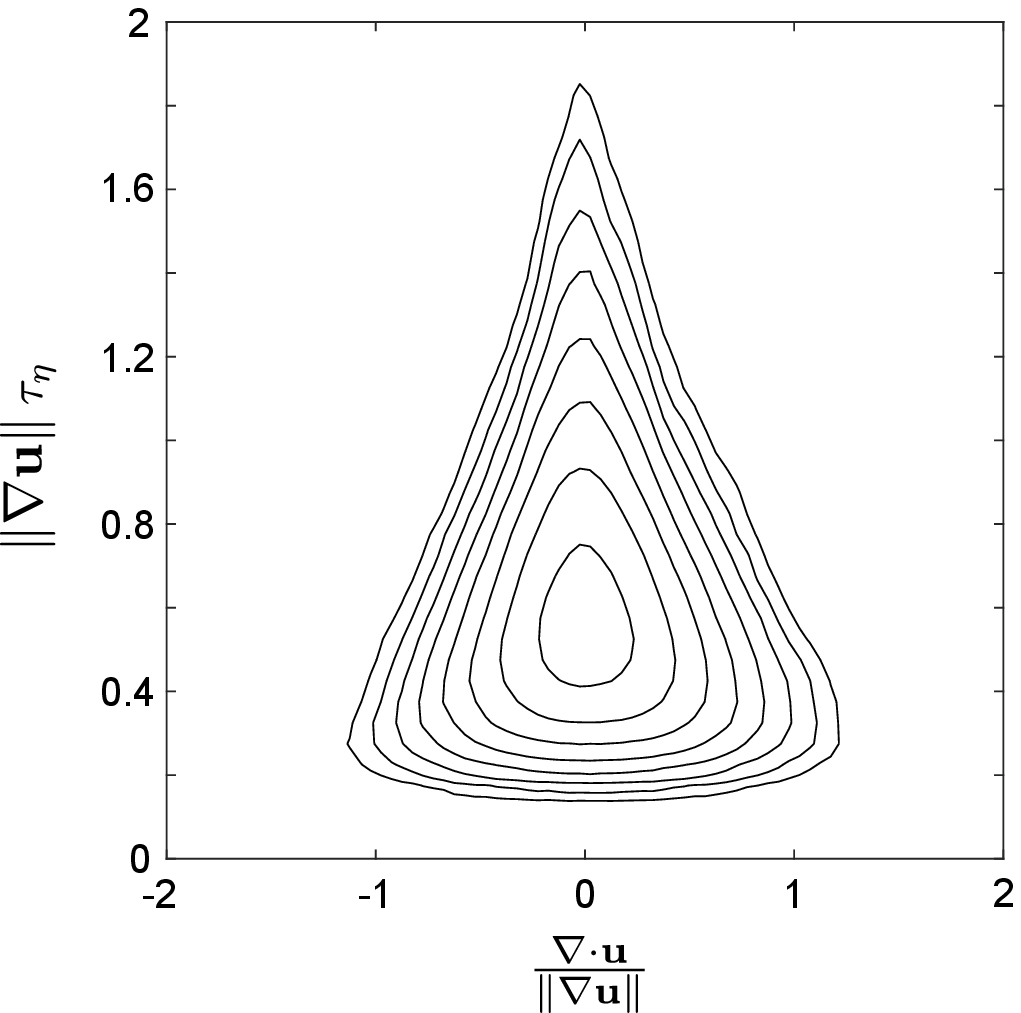}
\caption{Joint probability density function between the divergence relative to the module of the velocity gradient tensor and the module of the velocity gradient tensor itself non-dimensionalized by the Kolmogorov time scale $\tau_{\eta}$}. The contour levels are given on a logarithmic scale (base 10), in a range between -1.8 and 0.6, with increments of 0.3. 
\label{RelativeError}       
\end{figure}

\begin{figure*}[th]
  \includegraphics[width=0.97\textwidth]{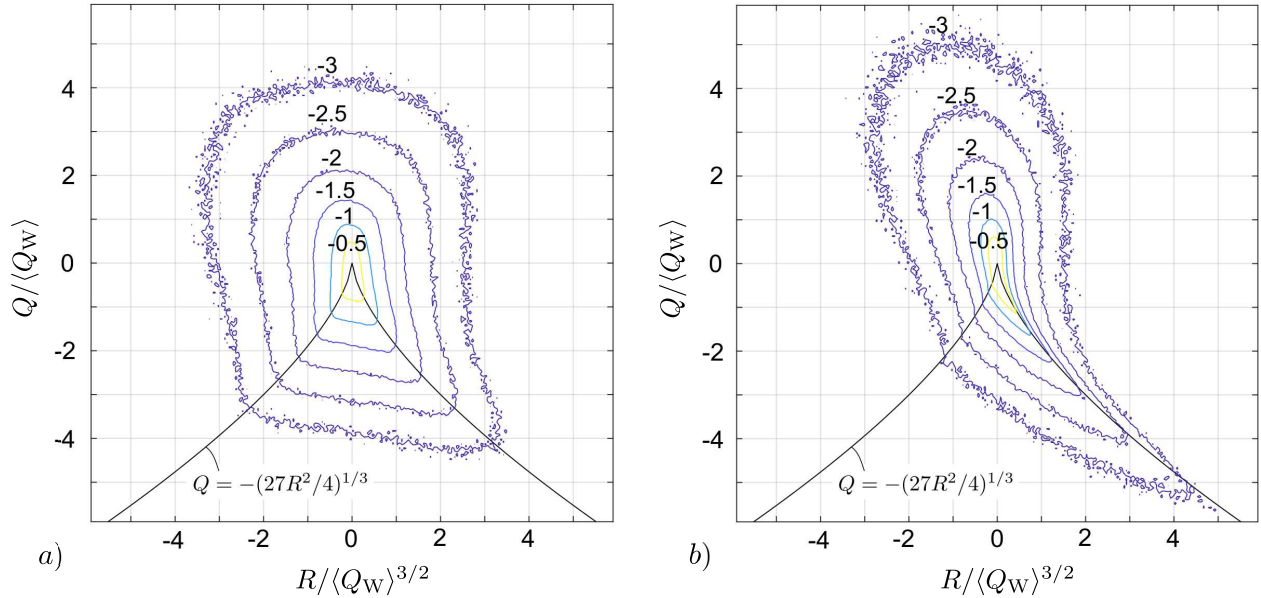}
\caption{Joint probability density functions of the second (Q) and third (R) invariants of the velocity gradient tensor obtained from velocity vector fields from a) the present experiment, and b) DNS simulations of homogeneous isotropic turbulence (HIT) at $Re_{\lambda} \approx 433$ \citep{Li2008}. The contour levels are on a logarithmic scale (base 10). The black continuous line is the null-discriminant line $D = Q + (25R^{2}/4)^{1/3} = 0$. $Q_\textup{W}=(\nabla \times \mathbf{u})^{2}/4$. }
\label{TD_both}       
\end{figure*}

\begin{figure}
  \includegraphics[width=0.45\textwidth]{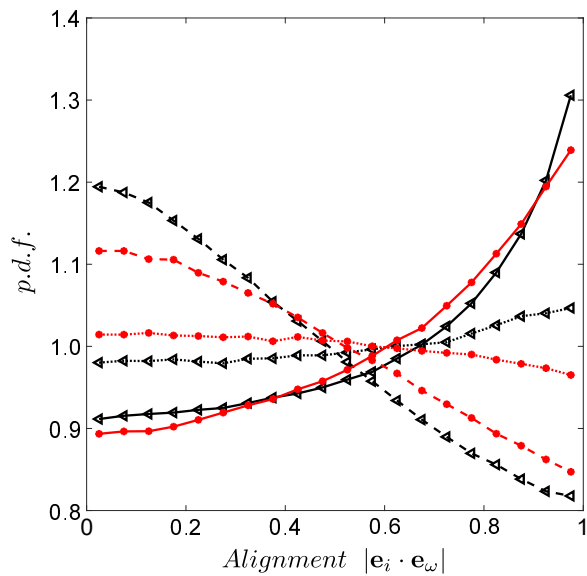}
\caption{Probability density functions of the cosine alignments between the eigenvectors of the strain-rate tensor and the vorticity vector. Red lines with circles are results from the present experiment, black lines with triangles are results from DNS simulations of homogeneous isotropic turbulence (HIT) at $Re_{\lambda} \approx 433$ \citep{Li2008}. Extensive, intermediate, and compressive eigenvectors are identified respectively by  dotted lines, dashed lines, and continuous lines. }
\label{alignment}       
\end{figure}

In the present section, we assess the results from tomographic long-range \textmu PIV by examining several turbulence quantities. A first important validation can be obtained from a comparison of the velocity statistics with previous experiments conducted on the same turbulent flow. Table \ref{statistics} presents the turbulence statistics as estimated from hot-wire anemometry and from long-range \textmu PIV \citep{Fiscaletti2014a}. 
   \begin{table*}[ht]
\begin{centering}
    \begin{tabular}{l c c c c }
    \toprule
  & Mean [ms$^{-1}$] & R.m.s. [ms$^{-1}$] & Skewness & Kurtosis   \\
  \midrule
  Hot-wire (centerline) & 10.56 & 2.63 & 0.08 & 2.85  \\
  \textmu PIV & 9.78 & 2.69  & 0.10 & 2.84  \\
  Tomo \textmu PIV with VIC+ & 10.57 & 2.78  & 0.08   & 2.70   \\
Tomo \textmu PIV without VIC+ & 10.65 & 2.83  & 0.11   & 2.72   \\
\bottomrule
    \end{tabular}
    \caption{Turbulence statistics calculated from hot-wire anemometry and from long-range \textmu PIV as in \cite{Fiscaletti2014b}, and from the experiment of tomographic \textmu PIV object of the present work. }
    \label{statistics}
    \end{centering}
\end{table*}
The mean of the streamwise velocity appears to be in good agreement with the results from hot-wire anemometry, while the lower value found through planar \textmu PIV can most probably be attributed to the measurement location being not exactly at the jet center-line. The r.m.s. of the streamwise velocity exhibits a slight overestimation when comparing the two \textmu PIV experiments with hot-wire anemometry. Moreover, larger values of the r.m.s. are obtained for tomographic \textmu PIV when compared to planar \textmu PIV, which can be attributed to higher levels of noise of the former. The skewness is nearly the same in all three experiments. Regarding to the kurtosis, tomographic \textmu PIV tends to underestimate it, which is indicative of a less sharp peak in the probability distribution of the streamwise velocity. \\
With the aim of determining which scales of turbulence and which velocity components are mostly affected by noise, velocity spectra are calculated. Velocity vector fields from tomographic \textmu PIV and VIC+ post-processing are analysed. The top panels of figure \ref{spectra} show velocity spectra of the streamwise, transversal, and out-of-plane components of the velocity along the three Cartesian axes. $\kappa_{1}$, $\kappa_{2}$, and $\kappa_{3}$ represent the wavenumbers in the streamwise, transversal, and out-of-plane components respectively, defined as $\kappa = 2\pi /\zeta$, where $\zeta$ is the wavelength. The present range of wavenumbers includes part of the inertial sub-range and the dissipative range. 
A further assessment of the accuracy of a volumetric measurement is obtained by considering the equation of mass conservation for an incompressible flow: 
\begin{equation}
    \frac{\partial u}{\partial x} + \frac{\partial v}{\partial y} + \frac{\partial w}{\partial z} = 0
    \label{eqn:mass_conservation}
\end{equation}
The lowest wavenumber for the out-of-plane direction ($\kappa_{3}$) is prescribed by the size of the measurement volume along that direction, which is approximately 910 \textmu m. All spectra are characterised by three different regions. At $\kappa \eta < 0.2$ the last part of the inertial subrange can be identified, where the spectra approximately decrease with a $-5/3$ slope. The interval $0.2 < \kappa \eta < 0.6$ is part of the dissipation region, showing a steeper energy decrease with the wavenumber. The observed deviation of the spectra from the $-5/3$ slope for $\kappa \eta \gtrapprox 0.2$ is consistent with several previous studies from the literature, see \cite{Pope2000}, Fig. 6.14. At $\kappa \eta > 0.6$, the effects of noise clearly manifest themselves, which is evident from the spectra becoming flatter. It is however worth highlighting that the three directions are not equally affected by noise. The out-of-plane spectra reveal the largest noise levels regardless of the velocity component, while the other two directions are almost equally affected by noise, with a mildly larger noise level along the transversal direction. Furthermore, the out-of-plane component $w$ shows the highest noise level in the out-of-plane spectrum ($\kappa_{3}$). These observations are reinforced from the analysis of the dissipation spectra, which are presented in the bottom panels of figure \ref{spectra}. The dissipation spectra have been obtained by multiplying the energy spectra by $\kappa_{i}^{2}$, which amplifies the noise at small scales (large wavenumbers). Here, in the range where the velocity spectra flatten in consequence of noise, $\kappa \eta > 0.6$, a sharp increase of the energy content is observed, consistent with the analysis of \cite{Ganapathisubramani2007}. In addition, the dissipation spectra of the streamwise and of the transversal velocity components, respectively $u$ and $v$, reveal a peak at $\kappa \eta \approx 0.1$, consistent with previous observations \citep{Pope2000}, demonstrating that the small-scale motions have been (largely) resolved for those velocity components. The location of the peak in the dissipation spectra of the out-of-plane velocity component is shifted towards $\kappa \eta \approx 0.2$, which suggests that the small-scale fluctuations in $w$ are affected by noise. The larger uncertainties in $w$ as compared to the uncertainties in the other directions are explained by the relatively small opening angle between cameras. \\
The assessment involves the computation of velocity gradients from the experimental dataset. With the aim of further attenuating the measurement noise associated with the calculation of the velocity gradients, a regression filter was applied to the velocity vector fields reconstructed with VIC+. The filter fits a second order a polynomial function in a $12 \times 12 \times 12$ neighbourhood around a point. The velocity gradients are then obtained from the coefficient of the fitted polynomial. Details on this regression filter are given by  \cite{Elsinga2010b}. \\
The joint probability density function (j.p.d.f) between $-(\partial u/\partial x)$ and $(\partial v / \partial y + \partial w/ \partial z)$ is presented in figure \ref{continuity}. Approximately eighteen million points contribute to the statistics. The thick black line represents the equation of mass conservation, Equation \ref{eqn:mass_conservation}. The pattern obtained from the j.p.d.f. is that of an inclined ellipse, where the major axis of the ellipse is almost aligned with the condition for mass conservation. The minor axis of the ellipse indicates the level of deviation from mass conservation which is associated with random error. 
Although not reported here, the major axis of the ellipse pattern obtained without any VIC processing presents a much larger inclination angle, which results in a strong misalignment with respect to mass conservation. This most probably arises from errors in the $w$ component of the velocity, which dominates. The regularization of the marching algorithm is very effective in correcting this divergence error by small modifications of the velocity field, with a significant effect on the velocity gradients. \\
The correlation coefficient between \\ $-(\partial u/\partial x)$ and $(\partial v / \partial y + \partial w/ \partial z)$, is 0.31 for the velocity gradients post-processed with the regression filter solely, while it is 0.71 for the velocity gradients treated with the VIC+ followed by the regression filter. The value of 0.71 is in reasonable agreement with results from previous experiments, specifically with 0.82 from time-resolved stereoscopic PIV by \cite{Ganapathisubramani2007}, with 0.84 from tomographic PIV by \cite{Jodai2016}, with 0.7 from multi-probe hot-wires by \cite{Tsinober1992}, while a scanning approach in a slow flow can achieve a value of 0.98 \citep{Lawson2014}.  \\
After having characterized the level of deviation from mass conservation in an absolute sense, it is of interest to quantify this deviation in relation to the module of the local velocity gradient tensor, to determine its relative weight. The p.d.f. of the divergence non-dimensionalized by the module of the local velocity gradient tensor was therefore computed. Even though not shown here, this analysis gives a distribution analogous to that from \cite{Ganapathisubramani2007}, their figure 9c. The r.m.s. of our p.d.f. is 0.3, which compares well with aforementioned PIV measurements of \cite{Ganapathisubramani2007} and with the dual-plane stereoscopic PIV of \cite{Mullin2006}, who reported values of the r.m.s. of 0.25 and 0.35, respectively. 
The described p.d.f., however, does not clarify which gradients are mostly contributing to the relative error of mass conservation. The relationship between the non-zero divergence and the local magnitude of the velocity gradient tensor can be assessed in further detail by the j.p.d.f. of the divergence non-dimensionalized by the local module of the velocity gradient tensor and the module of the velocity gradient tensor non-dimensionalized by the Kolmogorov time scale. The outcome of this statistical analysis is presented in figure \ref{RelativeError}. Velocity gradients at low magnitude are characterized by a large relative divergence, whereas stronger and more intermittent gradients are associated to significantly lower levels of relative divergence error. Analogous results were found by \cite{Ganapathisubramani2007} (see their figure 9d). 
These observations suggest that regions of strong gradients, i.e. regions populated by structures of intense vorticity and dissipation, are relatively accurately captured, while in regions defined `sleeping' after \cite{Carter2018}, i.e. those of low enstrophy and thus low velocity gradients, the velocity measurement is more affected by noise.\\ 
In the analysis of the small scales of turbulence, two features appear as universal. These are (\textit{i.}) the j.p.d.f. of the second and third invariants of the velocity gradient tensor presenting a teardrop shape, (\textit{ii.}) the preferential alignment of the vorticity vector with the eigenvector associated with the intermediate eigenvalue of the strain rate tensor \citep{Elsinga2010a}. These aspects are considered universal, and therefore common to all turbulent flows at conveniently large Reynolds numbers. The described statistical quantities are calculated from the present dataset from tomographic long-distance \textmu PIV, and compared with those from DNS presented in the literature.
The j.p.d.f. of the second (Q) and third (R) invariants of the velocity gradient tensor is shown in figure \ref{TD_both}a. As expected, the shape of the j.p.d.f. resembles that of a teardrop, and is analogous to the shape obtained from previous analyses, both numerical \citep{Ooi1999} and experimental \citep{Buxton2010}. The j.p.d.f. of the second and third invariants enables to quantify in a statistical sense the local flow topology. According to \cite{Chong1990}, four regions can be identified based on the sign of R and of the discriminant $D= R^{2}/4 + Q^{3}/27$, corresponding to four local flow topologies. Region I corresponds to stable focus/stretching ($R > 0$, $D > 0$), region II to unstable focus/stretching
($R < 0$, $D > 0$), region III to stable node/saddle/saddle ($R > 0$, $D < 0$), and region IV to unstable node/saddle/saddle ($R < 0$, $D < 0$). 
From the j.p.d.f. in figure \ref{TD_both}a, the flow topology of focus/stretching appears to be prevalent within the turbulent flow, with a dominance of stable (region I) over unstable focii (region II). The points of node/saddle/saddle, associated with $D < 0$ are most probably pertaining to an unstable region (IV) rather than to a stable one (region III), as the j.p.d.f. is skewed towards positive values of R. More generally, the j.p.d.f. sees larger probability values in the neighborhood of the positive branch of the null-discriminant black line, $D=0$, thus exhibiting the so called 'Vieillefosse tail' \citep{Vieillefosse1984}. 
The present PIV measurement was conducted at 70 nozzle diameters downstream from the jet nozzle, nearly at the centerline. At this downstream location, the turbulence can be assumed homogeneous and isotropic. To validate the statistical topological content of the far-field jet, we compare it with that of homogeneous isotropic turbulence (HIT) from DNS simulations by John Hopkins University \citep{Li2008}. The turbulent flow is characterized by a Reynolds number based on the Taylor microscale ($Re_{\lambda}$) of 433, thus similar to the Reynolds number of the present study, of approximately 350. It is however worth stressing that using a flow with analogous turbulence properties for this comparison is not strictly necessary, given that the shape of the j.p.d.f. of the second and third invariants is a universal property of turbulence, as previously explained. An additional and more significant motivation for using this dataset is represented by its widespread use among the turbulence community, particularly for validation purposes. The j.p.d.f. of the second (Q) and third (R) invariants calculated from this numerical dataset is presented in figure \ref{TD_both}b. 
As can be observed, the shape of this j.p.d.f. is more narrow and vertically stretched than the shape of the j.p.d.f. of figure \ref{TD_both}a. In particular, the j.p.d.f. obtained from the present experiment exhibits a much shorter Vieillefosse tail. 
The extent of the differences between the two j.p.d.f.s can be quantified by comparing the relative percentages associated to each flow topology. These are presented in table \ref{invariants_percentage}. Region I shows analogous percentages. In region III, similar percentages are obtained for the present experiment and for the simulations of HIT. The most significant discrepancies between the experimental dataset and the DNS can be found when comparing regions II and IV. As a consequence of the shorter Vieillefosse tail, the percentage of points belonging to region IV is much lower for the experimental analysis. This underestimation of points of unstable node/saddle/saddle (region IV) is partially compensated by an overestimation of points of unstable focii (region II). Therefore, in the experiment there is a transfer of data from the strain-dominated regions ($D<0$) to the rotationally dominated regions ($D>0$), consistent with \cite{Buxton2011a} and \cite{Naka2016}. The observed underestimation of points of unstable node/saddle/saddle and overestimation of points of stable node/saddle/saddle can be the result of measurement noise, particularly in the tomographic reconstruction process. \\
   \begin{table}[h]
\begin{centering}
    \begin{tabular}{l c c c c }
    \toprule
  & \splitcell{$R > 0$\\ $D > 0$} & \splitcell{$R < 0$\\ $D > 0$} & \splitcell{$R > 0$\\ $D < 0$} & \splitcell{$R < 0$\\ $D < 0$}   \\
  \midrule
 DNS  & 36 & 25 & 9 & 30  \\
 Tomo \textmu PIV & 36 & 33  & 14 & 17  \\
\bottomrule
    \end{tabular}
    \caption{Relative percentages associated to each flow topology from the analysis of the second and third invariants of the velocity gradient tensor. }
    \label{invariants_percentage}
    \end{centering}
\end{table}
The cosine alignment between the eigenvectors of the strain-rate tensor and the vorticity vector is also calculated from the present experimental dataset. Three p.d.f.s quantifying the alignment tendencies associated with the three eigenvectors are presented in figure \ref{alignment}, using red lines with circles. The same alignment probabilities are calculated from the dataset of HIT from DNS simulations by John Hopkins University, which is shown in figure \ref{alignment} by black lines with empty triangles. The intermediate eigenvector (continuous lines) exhibits the largest probability of being aligned with the vorticity vector. 
The compressive eigenvector (dashed lines) tends to be more preferentially aligned orthogonally with respect to the vorticity vector. An almost flat p.d.f. is found for the alignment tendencies between the extensive eigenvector and the vorticity vector (dotted lines). These observations are consistent with previous studies from the literature \citep{Ashurst1987,Kerr1987,Luthi2005,Hamlington2008}, which validates the experimental results. A direct comparison with the numerical dataset of HIT reveals that the present measurement tends to underestimate the probability of alignment associated both with the compressive and with the intermediate eigenvectors. This observation is consistent with \cite{Buxton2011a} (their figure 14(b,c)), who found that noise tends to 'flatten' the p.d.f.s. Regardless, the three p.d.f.s obtained from this experiment exhibit small discrepancies if compared with those from the finely-resolved HIT. Even though not shown here, the alignment tendencies between the eigenvectors of the strain-rate tensor and the vorticity vector are very similar to those reported in figure \ref{alignment} when computed from velocity fields without VIC+. This result evidences that even though VIC+ strongly attenuates the divergence error, some small-scale statistical aspects do not present any significant improvements of their accuracy. \\
 \section{Small-scale structures from instantaneous snapshots}
\label{sec:snapshots}
\begin{figure}[hb]
  \includegraphics[width=0.45\textwidth]{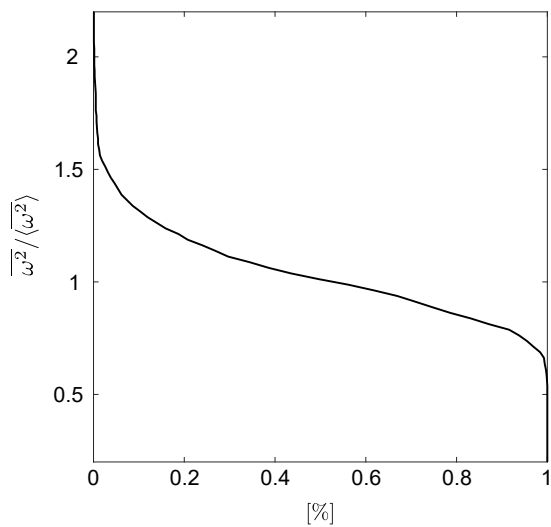}
\caption{Cumulative distribution function of the average non-dimensional enstrophy in each snapshot. }
\label{enstrophy}       
\end{figure}

\begin{figure*}[th]
  \includegraphics[width=1.0\textwidth]{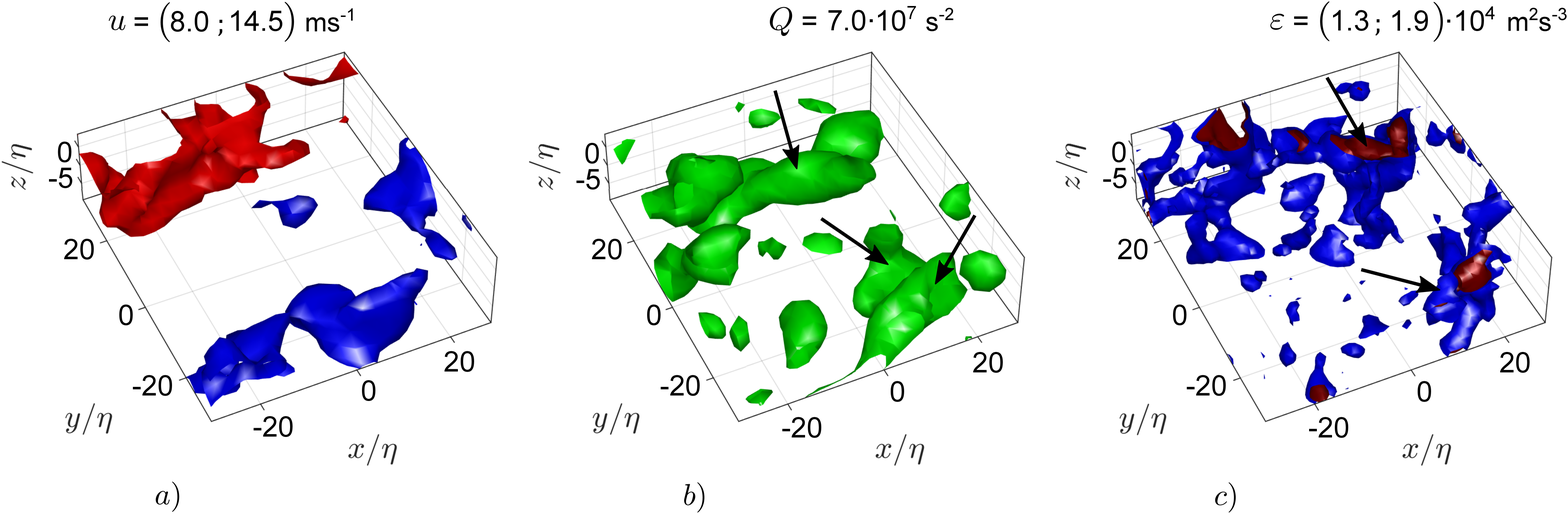}
\caption{Instantaneous snapshot from a so-called `hyperactive' state of the flow showing iso-surfaces of a) streamwise velocity, b) Q, and c) turbulent dissipation rate  $\varepsilon$. }
\label{structures}       
\end{figure*}

As discussed at the beginning of this section, the analysis presented in figure \ref{RelativeError} shows that events characterized by strong velocity gradients are affected by a mild relative divergence error. By leveraging the observation that the most extreme events are captured with a larger level of relative accuracy, we intend to examine the small-scale coherent structures appearing in the so-called `hyperactive' states of the flow \citep{Carter2018}, which are defined as instances when the spatially averaged enstrophy within the volume exceeds a certain threshold. The threshold is established using the cumulative distribution function of the average non-dimensional enstrophy (figure \ref{enstrophy}). 
The non-dimensionalization is made with respect to the total average enstrophy $\langle \overline{\omega^2} \rangle$, which is obtained by averaging over all realizations and the measurement volume. It can be seen that a very limited number of realizations, accounting for approximately $4 \%$ of the total sample, is characterized by a local enstrophy that is at least $50 \%$ larger than the total average enstrophy. This reveals that the hyperactive states are highly intermittent, consistent with previous observations from \cite{Ganapathisubramani2008}, \cite{Fiscaletti2014a}, \cite{Carter2018} among others. In the following, we consider a realization as hyperactive if $\overline{\omega^2} > 1.5 \langle \overline{\omega^2} \rangle$, similar to \cite{Carter2018} who adopted a threshold of 1.85. \\
The Q-criterion is then applied to the snapshots associated with hyperactive states with the aim of identifying the coherent structures of vorticity \citep{Hunt1988}. Similarly, the dissipation rate of turbulent kinetic energy is given in each point of the domain by: 
\begin{equation}
    \varepsilon = 2 \nu S_{ij} S_{ij}
\end{equation}
where $S_{ij}$ is the strain-rate tensor. An example of a snapshot reporting the spatial organization of the coherent structures is presented in figure \ref{structures}. The figure specifically shows, from left to right, iso-surfaces of the streamwise velocity, of Q, and of dissipation rate $\varepsilon$. When looking at figure \ref{structures}a, the instantaneous snapshot captures two zones of strongly different streamwise velocities, and the region of intense shear located in between these zones. The magnitude of this velocity difference is $\Delta u \approx 2.3 u_{rms}$, consistent with the velocity difference across significant shear layers as observed by  \cite{Hunt2014} and \cite{Ishihara2013}. In figure \ref{structures}b, coherent structures of vorticity tend to appear in the described shear region, consistent with the flow patterns from conditional averaging by \cite{Elsinga2010a}, with the instantaneous visualizations from DNS simulations by \cite{Ishihara2013}, and with the analyses by \cite{Hunt2014}.
\begin{figure*}[th]
  \includegraphics[width=1.0\textwidth]{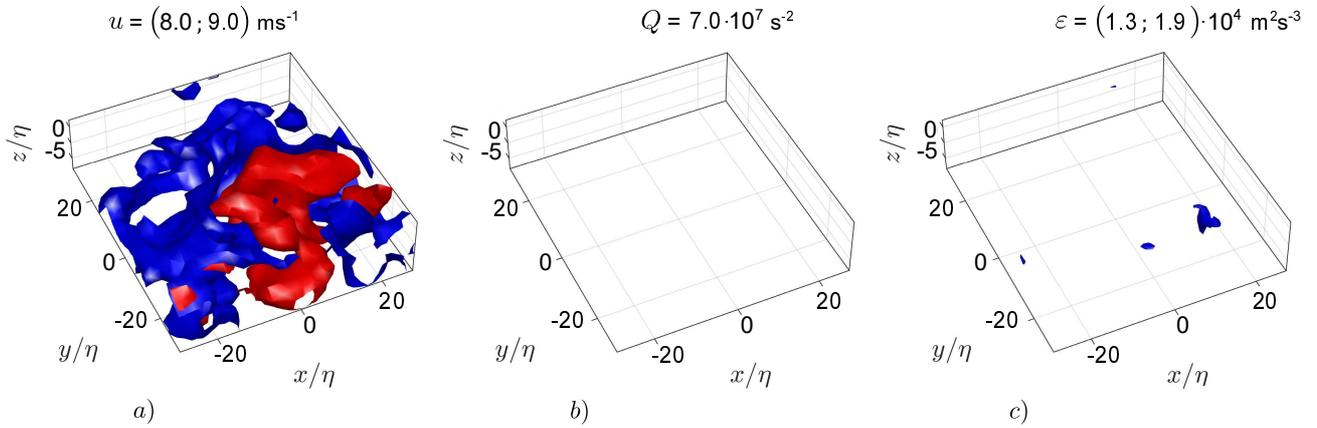}
\caption{Instantaneous snapshot from a so-called `sleeping' state of the flow showing iso-surfaces of a) streamwise velocity, b) Q, and c) turbulent dissipation rate $\varepsilon$. }
\label{sleeping}       
\end{figure*}
Several structures of intense vortices are clearly visible in the figure, three of which are highlighted by arrows. Even though the measurement volume is relatively small, the shape of the coherent structures of vorticity reminisces that of the small-scale worm-like vortices, or vortical tubes, found in numerous experimental and numerical investigations. On the other hand, a rather different shape characterizes the structures of intense dissipation, as can be seen in \ref{structures}c. These structures have a sheet-like appearance, which again agrees with the observations reported in previous studies, including \cite{Vincent1994} and \cite{Ganapathisubramani2008}. The structures of intense dissipation tend to be located in the vicinity of vortical tubes, therefore in regions of high shear. In figure \ref{structures}c, two examples of intense dissipation structures carrying the described features are highlighted. The levels of dissipation rates characterizing these intense structures is between 40 and 60 times larger than the average dissipation rate $\bar{\varepsilon}$ (see table \ref{flow_turbulence}). This is in fairly good agreement with \cite{Elsinga2020}, who found a ratio of about 70 between the maximum dissipation rate within regions of intense shear and the average dissipation rate of the flow, at $Re_{\lambda} \approx 350$. Overall, in support of the evidence that the described characteristics of the small-scale structures are common features to all hyperactive states and not simply a peculiarity of this snapshot solely, four additional snapshots were added to the Supplementary Material. \\
The analysis conducted up until here only focused on the hyperactive states of turbulence. However, the opposite extreme of the range reported in figure \ref{enstrophy} is characterized by very low levels of enstrophy, which led \cite{Carter2018} to define these states as `sleeping'. To better appreciate the difference between hyperactive and sleeping states, we also present an instantaneous realization where the local average enstrophy within the volume is lower than $70 \%$ of the total average enstrophy, i.e. $\overline{\omega^2} < 0.7 \langle \overline{\omega^2} \rangle$, therefore a realization associated with a sleeping state. This realization is given in figure \ref{sleeping}. From the iso-surfaces of $u$, the velocity within the measurement volume is rather uniform (figure \ref{sleeping}a), which leads to have weak levels of shear. As a result, no vorticity structures can be identified in the volume (figure \ref{sleeping}b). Analogous observations apply to the dissipation rate, shown in figure \ref{sleeping}c, where only one very tiny structure can be identified.  \\
The analysis of the extreme events of intense vorticity and of intense dissipation rate indicates that the experimental setup can confidently capture the spatial organization and the geometric characteristics of the small-scale structures within the turbulent flow under investigation. 

\section{Conclusions}
\label{sec:conclusions}
The present paper demonstrates that tomographic PIV can be extended to microscopic resolution at relatively large operating distance, which enables to measure high Reynolds number turbulence at laboratory scale. The novel experimental approach that was conceived has the aim of investigating the characteristics and the spatial organization of the small scales of turbulence in a jet at the Reynolds number based on the Taylor microscale ($Re_{\lambda}$) of 350, in which the Kolmogorov length scale is approximately 60 \textmu m. The described experimental setup configures itself as a tomographic long-range \textmu PIV setup. The novelty of this approach consisted in positioning piano-convex lenses on the optical path between each camera and the measurement object, which led to achieve a magnification factor of around 3 at an operating distance of approximately 500 mm. Furthermore, the setup is scalable, and it can be used with different lenses to achieve different magnification factors and working distances. The Scheimpflug condition was satisfied when mutually positioning the cameras, the lenses, and the measurement object, thus enabling to have the full PIV image on focus. The spatial resolution of the reconstructed velocity vector fields, intended as the size of the interrogation volume, was of $6\eta$, corresponding to a vector spacing of $1.5\eta$ when adopting a $75\%$ overlapping. From the analysis of mass conservation, the measurement was affected by noise in the reconstruction process, which was a-posteriori partly attenuated by applying the voxel-in-cell (VIC) algorithm \citep{Schneiders2014}. The analysis of the spectra evidenced that the reconstructed velocity fields remained affected by some measurement noise that could not be completely removed, particularly along the out-of-plane velocity component. To mitigate the effects of the noise on physical quantities involving velocity gradients, a regression filter with a kernel of size $12 \times 12 \times 12$ was additionally applied. A wide range of statistical aspects involving the small scales of turbulence was examined and favourably compared against results both from DNS simulations and previous works from the literature, which validated the measurement. \\
Coherent structures of intense vorticity and of intense dissipation rate were identified from instances representing the so-called hyperactive states of the flow, i.e. realizations where the spatially averaged enstrophy is larger than a threshold. The structures of vorticity were found in the shape of worm-like vortices, while the structures of dissipation tend to preferentially appear as sheet-like structures, consistent with several previous studies. The observed coherent structures tend to be located within regions of intense shear, which sit within jumps of streamwise velocity, $\Delta u$, that are between 1.5 and 2.5 times larger than $u_{rms}$. These results provide experimental evidence in support of recent observations from DNS simulations of homogeneous isotropic turbulence, e.g. \cite{Ishihara2013,Elsinga2020}.


%

\begin{acknowledgements}
D.F. was partly funded by the Marie Skłodowska-Curie Actions
of the European Union’s Horizon 2020 Program under the Grant
Agreement No. 895478 - ANACLETO. The authors would like to thank Prof. Christian K\"ahler for lending the Infinity K2 long-range microscope.
\end{acknowledgements}

%
%


\end{document}